# Scanning Gate Microscopy on Graphene: Charge Inhomogeneity and Extrinsic Doping

Romaneh Jalilian[1,2], Luis A. Jauregui[2,4], Gabriel Lopez[2,4], Jifa Tian[1,2], Caleb Roecker[3], Mehdi M. Yazdanpanah[5,6], Robert W. Cohn[6], Igor Jovanovic[3], Yong P. Chen[1,2,4,*]

[1] Department of Physics, Purdue University, West Lafayette, IN, 47907 USA
[2] Birck Nanotechnology Center, Purdue University, West Lafayette, IN, 47907 USA
[3] School of Nuclear Engineering, Purdue University, West Lafayette, IN, 47907 USA
[4] School of Electrical and Computer Engineering, Purdue University, West Lafayette, IN, 47907 USA
[5] NaugaNeedles LLC, Louisville, KY, 40299 USA
[6] Department of Electrical engineering, University of Louisville, Louisville, KY, 40292 USA
*Email: yongchen@purdue.edu

**Abstract**

We have performed scanning gate microscopy (SGM) on graphene field effect transistors (GFET), using a biased metallic nanowire coated with a dielectric layer as a contact mode tip and local top gate. Electrical transport through graphene at various back gate voltages is monitored as a function of tip voltage and tip position. Near the Dirac point, the dependence of graphene resistance on tip voltage shows a significant variation with tip position. SGM imaging reveals mesoscopic domains of electron-doped and hole-doped regions. Our measurements indicate a substantial spatial fluctuation (on the order of $10^{12}/cm^2$) in the carrier density in graphene due to extrinsic local doping. Important sources for such doping found in our samples include metal contacts, edges of graphene, structural defects, and resist residues.

The past few years have witnessed intensive research on graphene (2D carbon), a single layer graphite with unique electronic properties[1-3] and exciting promise for applications ranging from nanoelectronics to sensors. As a zero-gap semiconductor (semimetal), graphene's conduction and valence bands touch at the charge-neutral Dirac point (DP) with relativistic linear energy-momentum dispersion[1]. Intrinsically charge neutral, graphene can be easily doped electrically or chemically[3-6]. For example, in a typical graphene field effect transistor (GFET), a voltage applied to a gate (capacitively coupled to graphene) can tune the charge carriers (effectively tuning the Fermi energy relatively to DP) from p-type (holes) to n-type (electrons), with the graphene resistance peaking at the charge neutral DP[1,3]. Such an ambipolar electric field effect, which can exhibit high mobility even at room temperature, underlies the operation of most graphene devices. The *finite* minimum conductivity (which varies from sample to sample and shows discrepancies with earlier theories[1]) experimentally observed in graphene at DP has been a subject of much discussion. It is now understood that the values of minimum conductivity measured in realistic samples are largely related to charge inhomogeneity[7,8] in graphene, where the local charge density remains finite in the form of electron and hole puddles even while the *average* charge carrier density is zero at the (global) charge neutral DP. Various sources, such as topographic corrugations (e.g. ripples of graphene)[9], charged impurities[10,11], adsorbed molecules[5], surface contaminants[12], and metal contacts[6,13] have been suggested that could cause local doping and thus inhomogeneous charge density in graphene. The length scales of the resulting charge puddles and doping domains can range from nanometers (e.g. in the case of charged impurities[11]) to microns (e.g. in the case of metal contact[13]).

While several transport experiments have explored signatures of inhomogeneous charge density and doping in graphene[14-16], scanning probe microscopy (SPM) measurements provide the most direct probe of *local* electronic properties. Martin *et. al*. demonstrated the formation of sub-micron (resolution limited) electron and hole charge puddles near

DP in graphene using scanning single-electron transistor (SET) microscopy, and inferred the intrinsic size of the puddles to be ~30nm from measurements in the quantum Hall regime[8]. High-resolution scanning tunneling microscopy (STM) experiments have directly imaged charge puddles of ~10nm in size and suggested they originate from individual charged impurities underneath graphene[11,17]. Scanning photocurrent microscopy (SPCM) revealed modulations in the electrical potential across the graphene (particularly near electrodes) and gave evidence for metal-induced doping[18,19], although alternative interpretations of SPCM data in terms of photo thermal electric effects (PTE) have also been suggested[20].

In this paper, we report our study using atomic force microscope (AFM) based scanning gate microscopy (SGM) to probe the local electronic properties and charge inhomogeneity in graphene (exfoliated and supported on $SiO_2$). In SGM[21-29], a charged tip is used as a movable local top gate to modulate the carrier density underneath the tip in a device, whose resistance (or conductance) is measured. Scanning the tip (top gate) over the device generates a map showing how the electrical resistance (or conductance) through the device depends on the tip-induced local density modulation (local potential) at various locations. SGM has been previously applied to study the local electronic properties and defects in semiconductor nanostructures[21-24], nanowires[25] and carbon nanotubes (CNT)[26-29]. Similar techniques have recently been applied to graphene to study effects of local scattering potential on the quantum interference of carriers[30] and effects of current annealing on electronic uniformity[31]. Previous SGM measurements are typically performed in the "lift mode", where a biased conductive AFM tip is kept at a constant height (with possibly a small modulation[32]) above the device surface and the corresponding sample conductance response is measured. Sensitivity, spatial resolution (nonlocal effects of the tip), tip drift over time and measurement synchronization are some of the common challenges in the use of this technique[22,32]. In our work, we used an alternative method to perform SGM based on *contact-mode* AFM with a dielectric-coated metallic nanowire as the tip (Figure 1a-c). This scheme simplifies the SGM measurements with a number of technical advantages to be discussed below. Employing such a contact-mode SGM, we have obtained clear images demonstrating electronic inhomogeneity in graphene, particularly charge puddles formed near the DP.

The experimental setup of our contact-mode SGM is schematically shown in Figure 1a. One electrode ("drain") of the graphene device was grounded and DC bias voltages were applied to the tip (local top gate) and/or substrate (back gate). The graphene resistance ($R$) was measured at the room temperature (300K) and by passing a small source-drain current ($I_{ds}$=1 µA) while measuring the voltage drop ($V_{ds}$) cross the sample. We have performed 2-terminal, 3-terminal and 4-terminal measurements on various devices yielding qualitatively similar results for the purpose of this work. $V_{ds}$ can be fed into the AFM controller while scanning the tip to produce SGM images. Each SGM image has 512 lines, and each line (scanned at a rate of 0.68Hz) contains 512 sample points.

Contact mode AFM cantilever probes (spring constant 0.1 N/m) with metallic $Ag_2Ga$ nanowires (NWs) of high aspect ratio (50:1) at the end (HARNP-C20, NaugaNeedles, KY) were used for this study (Figure 1b&c). The flexible tip has a gentler contact with the surface that prevents scratching the graphene. The high aspect ratio and cylindrical structure of the NW reduce the parasitic capacitance between the tip (top gate) and the sample. The $Ag_2Ga$ NW is grown by immersion of the AFM tip (coated by a silver film, thickness 50-100 nm) in a small Ga droplet followed by slow extraction of the cantilever from the droplet[33,34]. NWs used in this work have typical diameter in the range of 50-100 nm and length of 1-5 μm. For the top gate dielectric, we used parylene-N (typical thickness in the range of 50-100nm, deposited in a thermal chemical vapor deposition system) conformally coated on the AFM tip with NW (Figure 1c). The uniform parylene coating facilitates a well-controlled dielectric thickness for the contact-mode scanning top gate that is also less susceptible to tip drift than conventional lift-mode SGM.

The monolayer graphene samples used in our experiments were fabricated by mechanical exfoliation[3] of highly ordered pyrolytic graphite (HOPG) on 300nm (thermally grown) $SiO_2$ on p-type doped Si substrate (back gate). The samples are fabricated into graphene field effect transistor (GFET) devices using e-beam lithography with evaporated Ti-Au (5nm-45nm) contact electrodes. Monolayer graphene is selected by optical microscopy and confirmed by Raman spectroscopy[35], and further verified by quantum Hall measurements[36,37] in selected devices. The mobility of our typical GFET devices is measured to be ~3000 $cm^2$/Vs.

Standard back-gated field effect measurements typically show a positive Dirac point (DP) in our graphene devices (Figure 1f), indicating extrinsic hole doping. Common sources for such doping include moisture from ambient environment and residues (eg. PMMA resist) from device fabrication processes. We have used regular contact-mode AFM to sweep away the dust and residues on the graphene surface and found that such AFM cleaning can reduce the extrinsic hole-doping (down-shift the Dirac point voltage), as demonstrated in Figure 1(d-f) for a device with a particularly high degree of residue coverage. Such cleaning is routinely performed for a more stable device response in subsequent contact-mode SGM. Topography of the graphene can be measured simultaneously during the contact-mode SGM, although regular tapping mode AFM is also performed to obtain topography images with slightly better quality.

The main results of this paper (scanning gate measurements) are presented in Figures 2-4. Multiple devices have been studied with qualitatively similar findings and representative data from 3 devices ("A","B" and "C") will be presented below. We first show the effect of local top gate (AFM tip) voltage on graphene resistance. The AFM image (tapping mode) of the device (sample A) used for this measurement is shown in Figure 2a. The graphene resistance ($R$) is measured between contacts 1 and 2. Figure 2b shows the "global" field effect by sweeping the back gate voltage ($V_{bg}$), which controls the global *average* charge carrier density ($\langle n \rangle$) in graphene. The global Dirac point voltage ($V_{DP}$) is 8.5V for this device. Figure 2c, d & e show $R$ measured as a function of top gate voltage ($V_{tg}$, swept from -20 to 20V) applied to the AFM tip for two different tip locations (marked in Figure 2a as L1 and L2) and at fixed back gate voltages of (c) $V_{bg}$=0V, (d) $V_{bg}$=8.5V and (e) $V_{bg}$=20V, respectively. In Figure 2c ($V_{bg} < V_{DP}$), the graphene is heavily p-type (with $\langle n \rangle \sim +6\times10^{11}$cm$^{-2}$, estimated from the global field effect[38] shown in Figure 2b) and $R$ increases with increasing $V_{tg}$ (within the range measured) for both tip locations. The opposite behavior is seen in Figure 2e ($V_{bg} > V_{DP}$), where the graphene is heavily n-type ($\langle n \rangle \sim -8 \times 10^{11}$cm$^{-2}$) and $R$ decreases with increasing $V_{tg}$ (within the measurement range). In Figure 2d ($V_{bg} = V_{DP}$), where the graphene is at its global charge neutral DP ($\langle n \rangle \sim 0$), the R-$V_{tg}$ curve is generally non-monotomic within the range of measurement and displays a peak, which we call the "local" Dirac point (LDP). Furthermore, the R-$V_{tg}$ dependence is found to be highly dependent on the tip locations. For example, for location L1, the LDP occurs at $V_{tg} \sim +3$V, while for location L2 the LDP occurs for $V_{tg} \sim -6$V. Such spatial variation of LDP is a result of charge inhomogeneity in the graphene sample, as will be further addressed in the following. As a consistency check to confirm the gating effect of the biased tip, we have retracted the tip far away from the graphene surface and observed $R$ becoming insensitive to the voltage and position of the tip.

Figure 3 presents the results of SGM imaging on a GFET (sample "B", with a global $V_{DP} \sim 9$V) measured at a constant $V_{tg}$ (20V) at various $V_{bg}$. The AFM tapping mode image of this device is shown in Figure 3a. The parylene coating on the AFM tip has a thickness of 100 nm in this measurement. Figure 3b, c & d display the SGM image (resistance of GFET as a function of tip position) for $V_{bg}$=5V, 9V and 12V respectively. In Figure 3b, where $V_{bg}$= 5V (<$V_{DP}$), placing the tip on the p-type graphene is seen to *increase* its resistance R (by as much as nearly 1 kΩ compared to the background value when the tip is far away from graphene), with the graphene appearing *blue* (indicating higher resistance than background) in the SGM image. This is due to the local reduction of carriers (holes) density in graphene under the positively biased tip. The opposite behavior is observed in Figure 3d, where $V_{bg}$= 12V (>$V_{DP}$) and placing the tip on the n-type graphene *decreases* R and makes it appear *red* (indicating lower resistance than background) in the SGM image. This is due to the local enhancement of carrier (electron) density in graphene under the positively biased tip. However, in Figure 3c, where $V_{bg}$=9V (close to $V_{DP}$) and the graphene is in its global "charge-neutral" ($\langle n \rangle$~0) state, the "polarity" of the resistance response of graphene to the tip becomes spatially non-uniform (R can be either increased or decreased depending on the locations of the tip on graphene). In the SGM image, this is manifested as the graphene appearing to break into several "islands" with very different colors. These "islands", irregularly shaped and with length scales ranging from ~0.5-2μm, will be interpreted as resulting from electron or hole "charge puddles" formed in the graphene near its global "charge-neutral" DP due to inhomogeneous extrinsic doping.

We have also studied how the SGM image showing the "puddles" develops with $V_{tg}$ while fixing $V_{bg}$ ~ $V_{DP}$. The results measured for Sample "C" are presented in Figure 4. The contact mode AFM image (acquired simultaneously with the SGM images) of this device is shown in Figure 4a. In this experiment, $V_{bg}$ is fixed at 14V, which is the measured global $V_{DP}$ for this device. The parylene coating on the SGM tip used has a thickness of 50nm. Figure 4b-g displays SGM images taken with $V_{tg}$ varying from 3V to -2V (in 1V step), plotted with the same color scale and resistance span (from 6.45 kΩ to 6.75 kΩ). The "puddle" pattern, qualitatively similar to that observed in Figure 3c and in all other samples we measured near the DP, is again observed in Figure 4b (with a positive $V_{tg}$= 3V). The pattern (and the magnitude of the spatial variation of resistance response to tip location) is seen to subdue (Figure 4c-e) as the tip voltage is reduced and almost disappears at $V_{tg}$=0V (Figure 4e)[39]. The "puddle" pattern in the SGM images is seen to re-appear for negative $V_{tg}$ (Figure 4f-g), but with reversed "polarity" (switching the blue and red regions, or enhanced-R or depressed-R regions) from positive $V_{tg}$.

It is known that, due to various sources of disorder and extrinsic doping, the carrier density in a realistic graphene sample and GFET device is spatially inhomogeneous. A biased SGM tip (top gate) can capacitively induce or deplete charge carriers in graphene. The main features of our observations can be understood simply by considering how the electronic transport of graphene with an inhomogeneous carrier density can be affected by local modulation of charge carriers due to the tip. We have used finite-element analysis (COMSOL) to simulate the electrostatic potential (*V*) generated by the biased contact-mode SGM tip. The result for a representative tip with NW diameter of 100nm, parylene thickness of 100nm and $V_{tg}$ of 1V is shown in Figure 5a. In our model, the tip is assumed to have radial symmetry with a "round" end (Figure 5a), a good approximation to the shape shown in the SEM image (Figure 1c). The graphene is modeled as an electrically grounded plane[40] touching the tip (parylene) at the end point. We have simulated the effect of small variations of the tip geometry and tip-graphene contact area and found the results do not qualitatively change our conclusions. The biased tip would deplete charge carriers (or induced charges with opposite polarity) in graphene underneath. The induced surface charge density, calculated from $\left.\frac{\partial V}{\partial y}\right|_{y \to 0+}$ (change of electric field normal to graphene), for the tip shown in Figure 5a is plotted (as a function of the radial distance in the graphene plane) in Figure 5b. The locally induced charge density decays away from the tip contact point with a characteristic length scale (full width at half maximum, FWHM) of ~130nm. Figure 5c schematically depicts a spatially-fluctuating carrier density $n(x)$ (thin solid blue line, excluding the tip-induced charges) and how a charged SGM tip may change the carrier density at various locations (thin dashed red line. The picture is drawn for a tip with $V_{tg}<0$ (the situation is simply reversed for $V_{tg}>0$). When graphene is globally p-type ($V_{bg}<<V_{DP}$ and $n(x)>>0$, with the thick dashed black line representing the zero carrier density level), a tip with $V_{tg}<0$ would decrease the graphene resistance ($R$) by adding charge carriers (holes) to the sample, whereas a tip with moderate[41] $V_{tg}>0$ would increase $R$ by depleting charge carriers. This is consistent with our observations in Figure 2c and 3b. The reverse is true when graphene is n-type ($V_{bg}>>V_{DP}$ and $n(x)<<0$, marked by the thick dot-dashed black line), consistent with our observations in Figure 2e and 3d. When graphene is near its (global) "charge-neutral" state ($V_{bg} \sim V_{DP}$), the *average* carrier density $<n> \sim 0$ (marked by the thick solid black line[42]). Because of the spatial fluctuation in $n(x)$, some regions of the sample have $n(x)>0$ (hole puddles) and some others have $n(x)<0$ (electron puddles). Now the response of $R$ would depend on the tip location. As displayed in Figure 5c, a tip with moderate $V_{tg}<0$ would decrease $R$ when placed above a hole puddle (e.g. location "1") and increase $R$ when placed above an electron puddle (e.g. location "3"), with reversed behavior for a tip with moderate $V_{tg}>0$. This allows us to identify the "red-shifted" (lower $R$) regions in graphene as electron puddles and "blue-shifted" (higher $R$) regions as hole puddles in SGM images taken with positively biased tips (e.g. Figure 3c, 4b, c), and reversely for SGM image taken with negatively biased tips (Figure 4f, g).

Applying this analysis to Figure 2d, we may associate location L1 with a hole puddle and L2 with an electron puddle from the respective response of $R$ at small $V_{tg}$. The LDP (maximum $R$ at finite $V_{tg}$) is understood because a tip that depletes local carriers at moderate bias could induce opposite-type carriers (e.g. location "2" depicted in Figure 5c) in graphene and lower $R$ with further increased tip bias. Therefore, the value of $V_{tg}$ at LDP (together with the calculated tip-induced charge density, Figure 5b) can be used to give an estimate for the local carrier density $n(x)$: $\sim +5\times10^{11}$ cm$^{-2}$ for L1 and $\sim -1\times10^{12}$ cm$^{-2}$ for L2. The *variation* ($V_{bg}$-independent) of carrier density between the locations, $\sim 1.5\times10^{12}$ cm$^{-2}$, corresponding to a variation of local DP or Fermi energy ($E_F = \hbar v_F \sqrt{\pi n}$) $\sim 80$ meV (taking the Fermi velocity[43] in graphene $v_F \sim 1\times10^6$ m/s), is comparable with the values obtained in other experimental[11,17] (low T) and theoretical works[44].

The length scale (~200nm-2μm) of the charge puddles we observe in the SGM images taken in our samples at the DP is comparable with the puddle sizes observed in SET[8] and SPC[18,19] measurements, but much larger than those (~10-20nm) observed in STM measurements[11,17]. This reflects both the resolution limit (~100-200nm) of our tips (similar to those in SPC and SET experiments) and the multiple length scales associated with the charge fluctuations in real graphene samples, as discussed below.

Charge density inhomogeneity (which leads to the formation of electron/hole puddles near DP) in graphene has been actively studied due to its importance for the electronic properties of graphene and the device performance. Various sources have been proposed to cause *extrinsic doping* resulting in charge inhomogeneity. Such sources include charged impurities near graphene[11,17], adsorbed molecules[5], surface contaminants (e.g. resist residues)[12], structural disorder in graphene (e.g. ripples[9]) and metal contacts[6,13]. Our SGM data reveal electron puddles near the contact electrodes (e.g. Figure 3c, Figure 4c) in our samples. This indicates that the substantial amount of Ti used in our contact causes n-type doping in graphene, consistent with earlier works[6,18]. The size of the contact-induced doping regions can reach micron scale (therefore not tip-resolution-limited), consistent with theoretical predictions[13] and SPCM and SGM measurements[18,19,31]. We also have often observed hole-puddles near edges of graphene (eg. Figure 4c, indicated by arrows, the extent to which they are observable varies from sample to sample). This suggests that edges, which are chemically more active than the bulk of graphene, tend to hole-dope the graphene, possibly due to the environmental molecules (eg. $H_2O$) bonded or adsorbed on the edge. In a different measurement involving sample C, we have also observed a hole-puddle formed around a scratch made in the graphene (Supplemental Figure S1). The observed width of such edge-induced puddles is comparable with the tip size, and is likely resolution limited. Our experiment is relatively insensitive to charge puddles of length scale <100nm, such as those associated with isolated impurities underneath graphene[11]. Furthermore, within the resolution of our experiment, we have not detected any correlation between the topography (height fluctuations, measured by regular AFM imaging) of a sample with the "charge puddle" pattern imaged by SGM near the DP, similar to the findings from STM experiments[11,17].

While extensively applied to 1D or quasi-1D samples, SGM for 2D conducting thin films is usually challenging[21-29,32,45] as the charges added or subtracted (by the SGM tip) from a small fraction of the sample area typically has only very weak effect on the global resistance of the whole sample. The situation for graphene is much better because of the significantly reduced density of states and charge screening[5,43,46]. For the SGM tip shown in Figure 5a,b, the total amount of the charges induced/depleted in graphene is calculated to be ~150 $e$ (for $V_{tg}$=1V). From the typical measured top gate local field effect (e.g. Figure 2b), we estimate a charge sensitivity of our GFET device (even with its moderate mobility) can reach ~30-50 m$\Omega$/$e$. This demonstrates the excellent potential of GFET as room temperature charge sensors.

The transport mean free path $l$ extracted from the carrier mobility is below ~50nm in our samples, shorter than both the SGM tip size (~100nm) and the device size (several μm). Therefore our measurements are performed in the diffusive transport regime. The observation that at the (global) DP, the resistance of graphene can be further increased by the SGM tip bias (eg. Figure 2d, Figure 3, Figure 4) demonstrates that the "minimal conductivity" measured in our graphene device at the DP is not universal[43], but dependent on the charge inhomogeneity in graphene, as has been pointed out previously[7,8,16]. SGM provides a simple and reversible way to modify such charge inhomogeneity and study its effect on electronic transport in graphene.

As an AFM-based technique, SGM has a number of advantages in probing local electronic properties of nanoelectronic devices. It can not only probe but also *manipulate* the local charge or potential profile and study the influence on the electronic transport through the operating device. SGM can be performed at a wide range of temperature, pressure and various ambient environments, and allows large area scan. It does not heat the sample as in SPCM[20] and does not have as stringent requirements on substrates as in STM. Our contact-mode technique presented here allows both topography and SGM images simultaneously obtained in one measurement. The high-aspect-ratio nanowire tip we employed reduces the parasitic capacitance between the conventional AFM tips with the sample, and can give improved spatial resolution. Further technical improvements may include using thinner nanowires or carbon nanotubes[47] as tips to further improve the spatial resolution, using tip voltage modulation and lock-in detection to improve the sensitivity, and performing SGM at low temperatures (where the carrier phase coherence length becomes large[48]) to study quantum transport[23,30] in graphene.

In conclusion, realistic graphene devices are subject to various extrinsic sources that locally dope the graphene, resulting in a spatially inhomogeneous charge density and formation of electron and hole puddles of various length scales at the global charge-neutral Dirac point. We have performed a contact-mode scanning gate microscopy on graphene and shown that metal contacts, graphene edges, and resist residues can be important sources of extrinsic doping. Our measurements can complement other forms of scanning probe microscopies to reveal the multiple origins of charge inhomogeneity in graphene and how such inhomogeneity can affect the electronic transport of graphene devices.

*Acknowledgements.* This work has been partially supported by National Science Foundation (ECCS#0833689), Department of Homeland Security (#2009-DN-077-ARI036-02) and by Nanoelectronic Research Initiative (NRI) via Midwest Institute for Nanoelectronics Discovery (MIND). We thank Tom Kopley, Ron Reifenberger, Leonid Rokhinson and Lishan Weng for helpful discussions and John Coy for technical assistance.

## References

[1] Geim, A.K.; Novoselov, K. S.; *Nature Mater.* **2007**, *6*, 183-191.

[2] Geim, A.K.; *Science* **2009**, *324*(5934), 1530-1534.

[3] Novoselov, K.S.; Geim, A. K.; Morozov, S. V.; Jiang, D.; Zhang, Y.; Dubonos, S.V.; Grigorieva, I. V.; Firsov, A. A.; *Science* **2004**, *306* (5696), 666–669.

[4] Wang, X.; Li, X.; Zhang, L.; Yoon, Y.; Weber, P.K.; Wang, H.; Guo, J.; Dai, H.; *Science* **2009**, *324*(5928), 768 – 771.

[5] Wehling, T.O.; Novoselov, K.S.; Morozov, S.V.; Vdovin, E.E.; Katsnelson, M.I.; Geim, A.K.; Lichtenstein, A.I.; *Nano Lett*. **2008**, *8* (1), 173-177.

[6] Giovannetti, G.; Khomyakov, P.A.; Brocks, G., Karpan,V.M.; Brink, V.D.; Kelly, P.J. *Phys. Rev. Lett*. **2008**, *101*, 026803.

[7] Adam, S.; Hwang, E. H.; Galitski, V.M.; Das Sarma, S.; *Proc. Natl. Acad. Sci. USA* **2007**, *104* (47), 18392-18397.

[8] Martin, J.; Akerman, N.; Ulbrich, G.; Lohmann, T.; Smet, J. H.; Von Klitzing, K.; Yacoby, A.; *Nature Phys.* **2008**, *4* (2), 145-148.

[9] Kim, E.A.; Castro Neto, A.N.; *Europhys. Lett.* **2008**, 84, 57007.

[10] Chen, J.H.; Jang, C.; Adam, S.; Fuhrer, M. S.; Williams, E. D.; Ishigami, M.; *Nature Phys.*, **2008**, *4* (5), 377 – 381.

[11] Zhang, Y.; Brar, V.W.; Girit, C.; Zett, A.; Crommie, M.F.; *Nature Phys.* **2009**, *5* (9), 722 – 726.

[12] Dan, Y.; Lu, Y.; Kybert, N.; Luo, Z.; Johnson, A.T.; *Nano Lett*. **2009**, *9* (4), 1472–1475.

[13] Golizadeh-Mojarad, R.; Datta, S.; *Phys. Rev. B*. **2009**, *79*, 085410.

[14] Cho, S.; Fuhrer, M. S.; *Phys. Rev. B.* **2008**, *77*, 084102.

[15] Huard, B.; Stander, N.; Sulpizio, J.A.; Goldhaber-Gordon, D.; *Phys. Rev. B* **2008**, *78*, 121402.

[16] Blake, P.; Wang, R.; Morozov, S. V.; Schedin, F.; Ponomarenko, L. A.; Zhukov, A. A.; Nair, R. R.; Grigorieva, I. V.; Novoselov, K. S.; Geim, A. K.; *Solid State Comm.* **2009**, *149,* 1068-1071.

[17] Deshpande, A.; Bao, W.; Miao,F.; Lau, C. N.; LeRoy, B. J.; *Phys. Rev. B* **2009**, *79*, 205411.

[18] Lee, E. J. H.; Balasubramanian, K.; Weitz, R.T.; Burghard, M.; Kern, K.; *Nature Nanotech.* **2008**, *3* (8), 486-490.

[19] Mueller, T.; Xia, F.; Freitag, M.; Tsang, J.; Avouris, P.; *Phys. Rev. B* **2009**, *79*, 245430.

[20] Xu, X.; Gabor, N. M.; Alden, J. S.; van der Zande, A. M.; McEuen, P. L.; *Nano Lett.* **2010**, *10* (2), 562–566.

[21] Eriksson, M.A.; Beck, R.G.; Topinka, M.; Katine, J.A.; Westervelt, R.M.; Campman, K.L.; Gossard, A.C.; *Appl. Phys. Lett.* **1996**, *69*, 671-673.

[22] Eriksson, M.A.; Beck, R.G.; Topinka, M.; Katine, J.A.; Westervelt, R.M.; Campman, K.L.; Gossard, A.C.; *Superlattices Microstruct.,* **1996**, *20,* 435-440.

[23] Topinka, M.A.; LeRoy, B.J.; Westervelt, R.M.; Shaw, S.E.J.; Fleischmann, R.; Heller, E.J.; Maranowski, K.D.; Gossard, A.C.; *Nature* **2001**, *410* (6825), 183-186.

[24] Aoki, N.; da Cunha, C. R.; Akis, R.; Ferry, D. K.; Ochiai, Y.; *Phys. Rev. B* **2005**, *72*, 155327

[25] Yang, C.; Zhong, Z.; Lieber, C. M.; *Science* **2005**, *310*, 1304-1307.

[26] Bachtold, A.; Fuhrer, M. S.; Plyasunov, S.; Forero, M.; Anderson, E. H.; Zettl, A.; McEuen, P. L.; *Phys. Rev. Lett*. **2000**, *84*, 266082.

[27] Tans, S.J.; Dekker, C.; *Nature* **2000**, *404*, 834-835.

[28] Bockrath, M; Liang, W.; Bozovic, D.; Hafner, J. H.; Lieber, C. M.; Tinkham, M.; Park, H.; *Science* **2001**, *291* (5502), 283-285.

[29] Freitag, M.; Johnson, A.T.; Kalinin, S.V.; Bonnell, D.A.; *Phys. Rev. Lett*. **2002**, *89*, 216801.

[30] Berezovsky, J.; Westervelt, R. M.; arXiv:0907.0428v1 (2009).

[31] Connolly, M. R.; Chiou, K. L.; Smith, C. G.; Anderson, D.; Jones, G.A. C.; Lombardo, A.; Fasoli, A.; Ferrari, A. C.; *Appl. Phys. Lett.* **2010**, *96*, 113501.

[32] Wilson, R. N.; Cobden, D. H.; *Nano Lett*. **2008**, *8* (8), 2161- 2165.

[33] Yazdanpanah, M. M.; Harfenist, S. A.; Safir, A.;Cohn, R. W.; *J. Appl. Phys.* **2005**, *98* (7), 073510.

[34] Yazdanpanah, M.M.; "Near room temperature self-assembly of nanostructure through Gallium reaction with metal thin films", PhD. Dissertation, University of Louisville (2006).

[35] Ferrari, A. C.; Meyer, J. C.; Scardaci, V.; Casiraghi, C.; Lazzeri, M.; Mauri, F.; Roth, S.; Geim, A.K; *Phys. Rev.Lett.,* **2006**, *97*, 187401.

[36] Novoselov, K.S.; Geim, A.K.; Morozov, S.V.; Jiang, D.; Katsnelson, M. I.; Grigorieva, I.V.; Dubonos, S.V.; Firsov, A.A.; *Nature* **2005**, *438*, 197-200.

[37] Zhang, Y.; Tan, Y.W.; Stormer, H. L.; Kim, P.; *Nature* **2005**, *438*, 201-204.

[38] We have $\langle n \rangle = C(V_{DP} - V_{bg})$, where C=$1.15 \times 10^{-4}$ C/Vm$^2$ is the capacitance per unit area between the graphene and the back gate.

[39] The small but finite contrast in SGM images even for $V_{tg} = 0$V may be due to the fringe electric field between the tip and the biased back gate[8], and the work function difference between the tip and graphene[30,49].

[40] Effects of the small source-drain voltages (~1 mV, much smaller than $V_{tg}$) used in our experiment can be neglected.

[41] Until $|V_{tg}|$ is so large to induce the opposite-type carriers in graphene and decrease its R. However, excessive $|V_{tg}|$ is typically avoided in our measurements to prevent breakdown of the parylene coating.

[42] Alternatively, one can view the thick black lines as the global chemical potential (Fermi level) set by $V_{bg}$ and the thin line as locus of Dirac point (local potential) to obtain similar conclusions.

[43] Castro Neto, A.H.; Guinea, F.; Peres, N.M.R.; Novoselov, K.S.; Geim, A.K.; *Rev. Mod. Phys.* **2009**, *81*, 109–162.

[44] Rossi, E; Das Sarma, S.; *Phys. Rev. Lett.* **2008**, *101*, 166803.

[45] Aoki, N.; Sudou, K.;Matsusaki, K.; Okamoto, K.; Ochiai, Y.; *J. Phys. Conf. Ser*. **2008**, *109*, 012007.

[46] Schedin, F.; Geim, A. K.; Morozov, S.V.; Hill, E.W.; Blake, P.; Katsnelson, M.I.; Novoselov, K.S.; *Nature Mater.* **2007**, *6*, 652-655.

[47] Wilson, N.R.; MacPherson, J.V.; *Nature Nanotech.* **2009**, *4*, 483–491.

[48] Morozov, S. V.; Novoselov, K. S.; Katsnelson, M. I.; Schedin, F.; Ponomarenko, L. A.; Jiang, D.;Geim, A. K.; *Phys. Rev. Lett.* **2006**, *97*, 016801.

[49] Yu, Y.J.; Zhao, Y.; Ryu, S.; Brus, L.E.; Kim, K.S.; Kim, P.; *Nano Lett.* **2009**, **9** (10), 3430–3434.

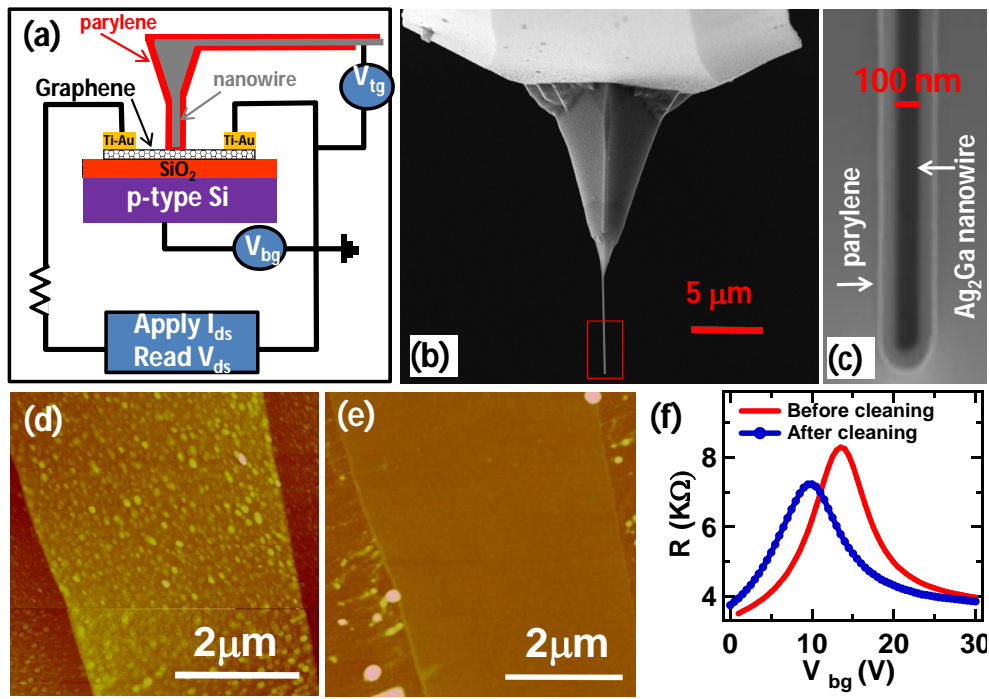

Figure 1. (a) Schematic of the experimental set up for contact mode scanning gate microscopy (SGM) on graphene. (b) SEM image of a custom-made contact mode SGM tip. (c) Magnified view of the end of the tip, showing a conductive $Ag_2Ga$ nanowire surrounded by parylene (dielectric) coating. (d) AFM image (tapping mode) of a graphene covered by residues from the device fabrication process. (e) Image of the same device in (d) after "nano-broom" cleaning by contact mode AFM. (f) The field effect (resistance vs. back gate voltage) of the GFET device before (d) and after (e) AFM cleaning.

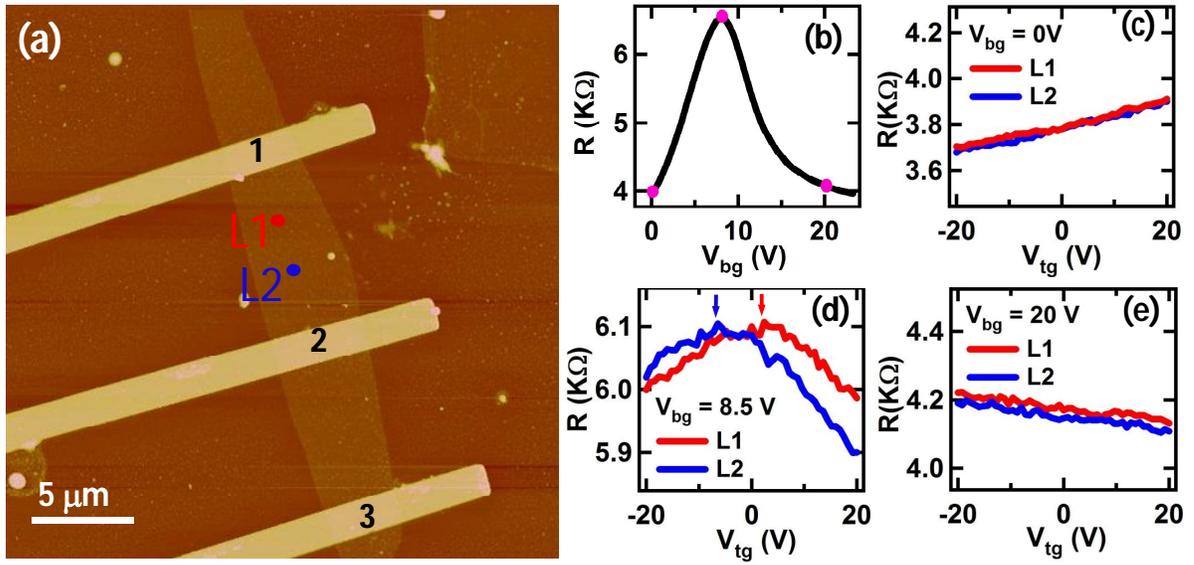

Figure 2. Global and local field effect (sample "A"). (a) AFM image (tapping mode) of the device. The graphene resistance ($R$) is measured between contact electrodes "1" and "2". (b) The "global" field effect: $R$ as a function of global back gate voltage ($V_{bg}$). The "global" Dirac point (DP) occurs at $V_{DP}$ ~8.5 V. (c-e) Local field effect: $R$ as a function of local top gate voltage ($V_{tg}$, applied to the SGM tip) measured at 3 different back gate voltages: (c) 0V ($V_{bg}<V_{DP}$), (d) 8.5 V ($V_{bg}=V_{DP}$) and (e) 20V ($V_{bg}>V_{DP}$). Data measured at two different tip locations (L1 and L2, marked in (a)) are shown in each panel (c-e). The thickness of the parylene coating on the AFM tip used is 100 nm.

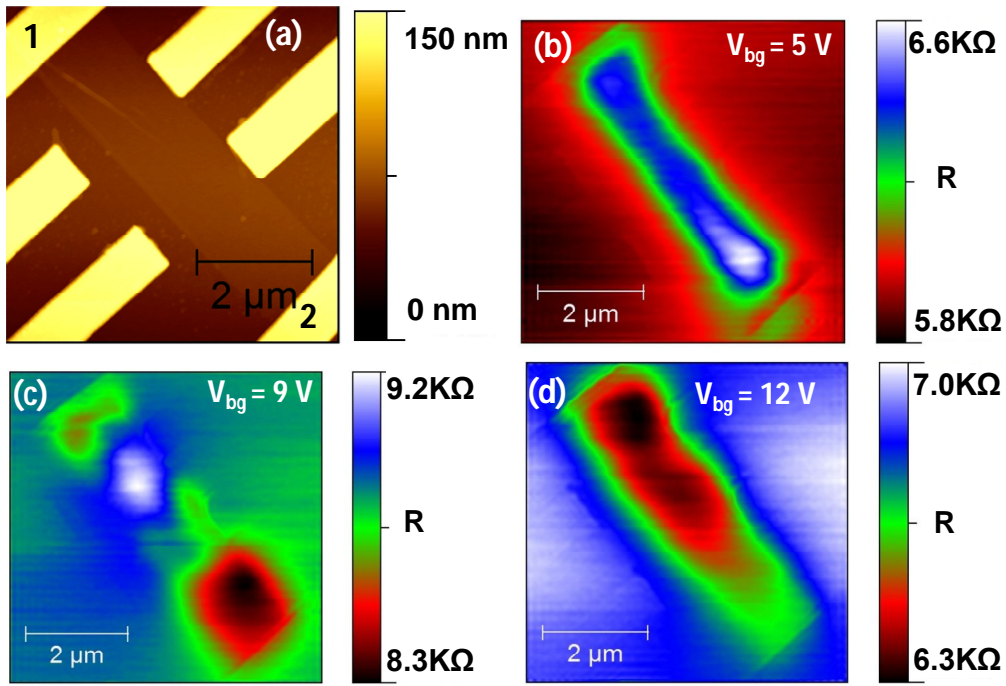

Figure 3. SGM at different back gate voltages (sample "B"). (a) AFM image (tapping mode) of the device. The graphene resistance (R) was measured between the contact electrodes 1 and 2 (the other electrodes shown are for a different experiment and kept floating in the SGM measurment). (b-d) SGM image of the GFET measured at 3 different back gate voltages: (b) 5V ($V_{bg}<V_{DP}$), (c) 9V ($V_{bg}=V_{DP}$), and (d) 12V ($V_{bg}>V_{DP}$). SGM imaging in (b-d) was performed at a fixed tip voltage ($V_{tg}=20V$) and over the same sample area shown in (a). The thickness of the parylene coating on the tip used is 100 nm.

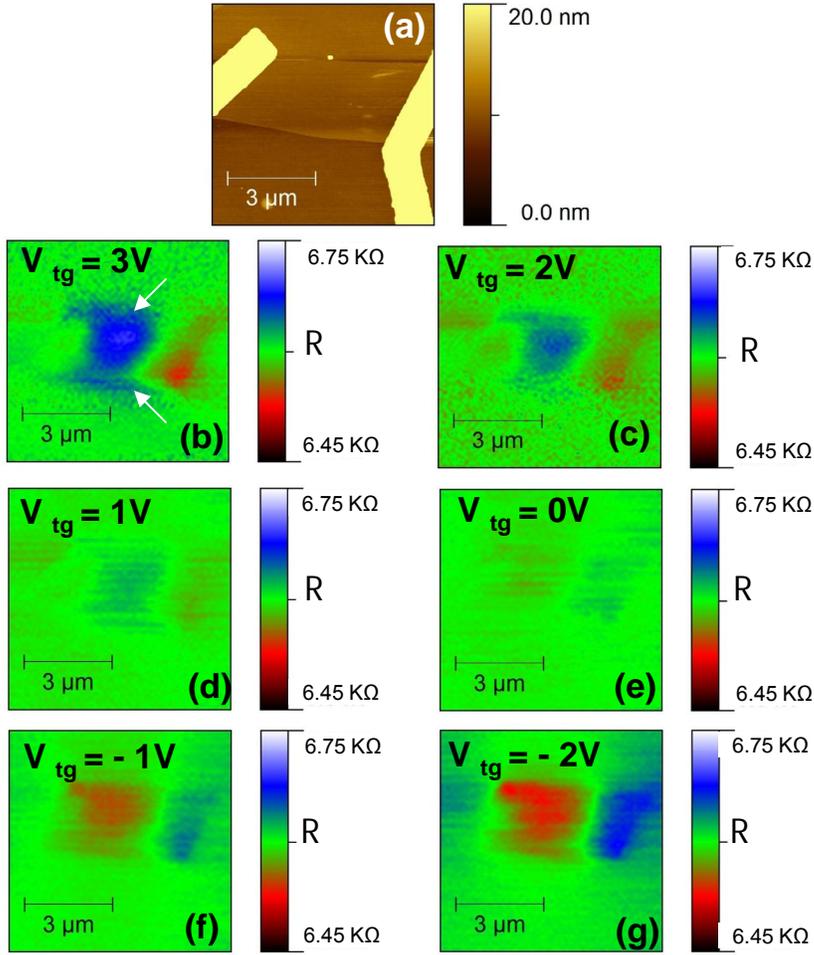

Figure 4. SGM at the global Dirac point at different top gate voltages (sample "C"). (a) AFM image (contact mode) of the device. The global Dirac point $V_{DP} = 14V$. (b-g) SGM over the same area shown in (a) taken at the DP ($V_{bg}=V_{DP}=14V$) with different tip voltages ($V_{tg}$ = 3, 2, 1, 0, -1, -2 V respectively). The same color scale (spanning ~0.3KΩ) is used for all SGM images. The white arrows in (b) mark two stripe-shaped "hole" puddles observed near and parallel to the edges of this graphene sample. The thickness of the parylene coating on the tip used is 50 nm.

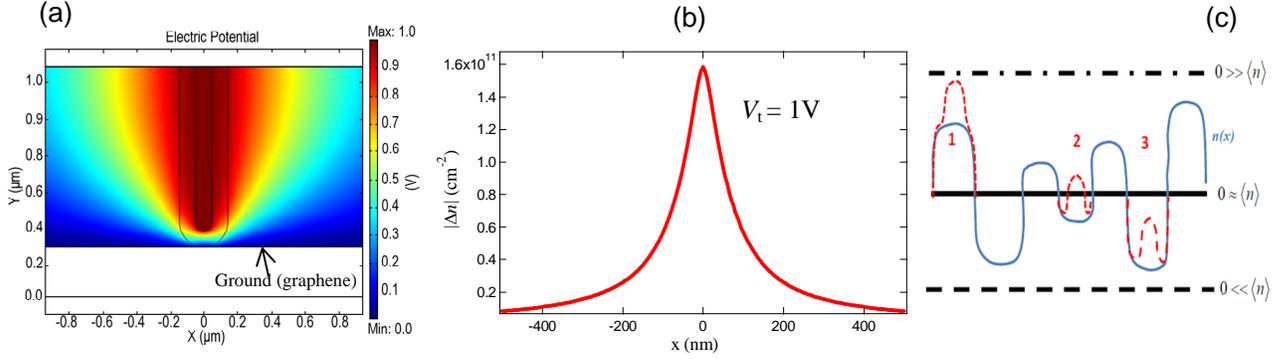

Figure 5. (a) Calculated electric potential of a representative nanowire-based SGM tip (with bias $V_{tg}=1V$) in contact with grounded graphene (indicated by arrow). The thickness of parylene coating separating the NW and graphene is 100 nm. The geometry is assumed to be radially (x) symmetric and the profile shown is a 2D cross section through the central axis (x=0) of the NW. (b) Calculated profile of charge density induced by the tip as shown in (a). (c) A schematic example of spatially inhomogeneous charge density $n(x)$ (thin blue solid line) and the change (thin red dashed line) due to a negatively biased SGM tip at 3 representative locations (labeled 1, 2 and 3). The thick black solid line, dot-dashed line and dashed line indicate the zero charge density level for three situations: charge neutral ($\langle n \rangle \sim 0$, with $V_{bg} \sim V_D$), n-type ($\langle n \rangle \ll 0$, with $V_{bg} > V_D$), and p-type ($\langle n \rangle \gg 0$, with $V_{bg} < V_D$) doping, respectively.

**Supplemental File**

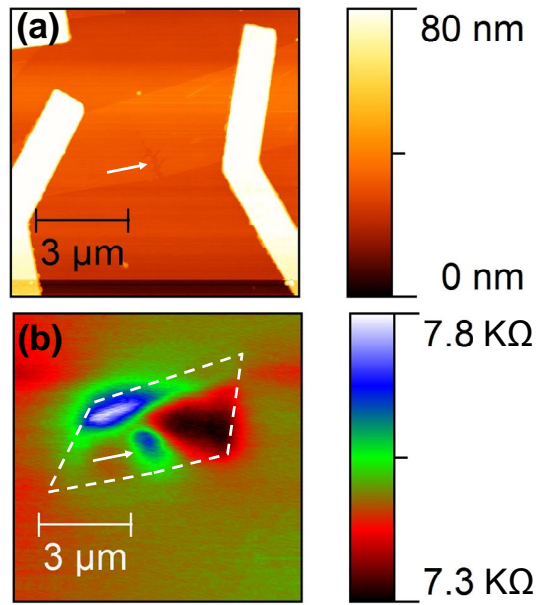

Figure S1. (a) AFM image (contact mode) of sample C (main text figure 4) after a scratch (indicated by arrow) was made by the AFM tip. (b) SGM image of the scratched sample biased near its global Dirac point. A hole-puddle was observed around the scratch (arrow). $V_{tg}$=5V was used in this measurement.